\documentclass[reprint,amsmath,amssymb,aps,prl,twocolumn,superscriptaddress,showpacs]{revtex4-1}

\usepackage{graphicx}
\usepackage{dcolumn}
\usepackage{bm}
\usepackage{color}
\usepackage{natbib}
\usepackage{soul}

\newcommand{\rev}[1]{\textcolor{black}{#1}}

\begin{document}
	
	\title{\rev{Uncovering the triplet ground state of triangular graphene nanoflakes \\ engineered with atomic precision on a metal surface}}
	
	\author{Jingcheng Li}
	\affiliation{CIC nanoGUNE, 20018 Donostia-San Sebasti\'an, Spain}
	\author{Sofia Sanz}
	\affiliation{Donostia International Physics Center (DIPC), 20018 Donostia-San Sebasti\'an, Spain}
	
	\author{Jesus Castro-Esteban}
	\affiliation{Centro Singular de Investigaci\'on en Qu\'imica Biol\'oxica e Materiais Moleculares (CiQUS), and Departamento de Qu\'imica Org\'anica, Universidade de Santiago de Compostela, Spain}
	
	\author{Manuel Vilas-Varela} 
	\affiliation{Centro Singular de Investigaci\'on en Qu\'imica Biol\'oxica e Materiais Moleculares (CiQUS), and Departamento de Qu\'imica Org\'anica, Universidade de Santiago de Compostela, Spain}
	
	\author{Niklas Friedrich} 
	\affiliation{CIC nanoGUNE, 20018 Donostia-San Sebasti\'an, Spain}
	
	\author{Thomas Frederiksen}   \email{thomas$\_$frederiksen@ehu.eus}
	\affiliation{Donostia International Physics Center (DIPC), 20018 Donostia-San Sebasti\'an, Spain}
	\affiliation{Ikerbasque, Basque Foundation for Science, 48013 Bilbao, Spain}

	\author{Diego Pe{\~{n}}a}   \email{diego.pena@usc.es}
	\affiliation{Centro Singular de Investigaci\'on en Qu\'imica Biol\'oxica e Materiais Moleculares (CiQUS), and Departamento de Qu\'imica Org\'anica, Universidade de Santiago de Compostela, Spain}

	\author{Jose Ignacio Pascual} \email{ji.pascual@nanogune.eu}
	\affiliation{CIC nanoGUNE, 20018 Donostia-San Sebasti\'an, Spain}
	\affiliation{Ikerbasque, Basque Foundation for Science, 48013 Bilbao, Spain}
	
\date{\today}

\begin{abstract}
Graphene can develop large magnetic moments in custom crafted open-shell nanostructures such as triangulene, a triangular piece of graphene with zigzag edges. 
Current methods of engineering graphene nano-systems on surfaces succeeded in producing atomically precise open-shell structures, but demonstration of their net spin remains elusive to date. Here, we fabricate triangulene-like graphene systems and demonstrate that they possess a spin $S=1$ ground state. Scanning tunnelling spectroscopy identifies the fingerprint of an underscreened $S=1$ Kondo state on \rev{these} flakes at low temperatures, signaling the dominant ferromagnetic interactions between two spins. Combined with simulations based on the meanfield Hubbard model, we show that this $S=1$ $\pi$-paramagnetism is robust, and can be manipulated to a  $S=1/2$ state by adding additional H-atoms to the radical sites. \rev{Our results demonstrate that $\pi$-paramagnetism of high-spin graphene flakes can survive on surfaces, opening the door to  study the quantum behaviour of interacting $\pi$-spins in graphene systems. }
\end{abstract}

\maketitle

In spite of their apparent simplicity, custom-crafted graphene nanoflakes (GNF) are predicted to exhibit complex magnetic phenomenology   \cite{Yazyev2010} with promising possibilities as  active components of a new generation of nanoscale devices  \cite{wang2009,Han2014,Bullard2015}. As predicted by Lieb's theorem for bipartite lattices  \cite{Lieb1989}, certain shapes of graphene structures may accommodate a spin imbalance in the $\pi$ electron cloud, resulting in GNFs with a net magnetic moment. Graphene $\pi$-paramagnetism is more delocalized, mobile, and isotropic than conventional magnetism from $d$ or $f$ states  \cite{Trauzettel2007}, and can be electrically addressable \cite{Li2019,Mishra2020}. Furthermore, the magnetic moments and their correlations in GNFs can be precisely engineered through their sizes, edge topology, or chemical doping  \cite{Fernandez-Rossier2007,Wang2008,Kan2012,maria2019}. 

The fabrication of such  GNFs has been hindered due to their high reactivity  \cite{Clar1953}. As open-shell structures, the presence of unpaired electrons (radicals) makes the synthesis  difficult. First unsubstituted triangulene was synthesized by dehydrogenating precursor molecules with the tip of a scanning tunneling microscope (STM)  \cite{Pavlicek2017}. Very recently, triangular GNFs with larger sizes have been synthesized through \rev{an} on-surface synthetic (OSS) approach \cite{Mishra2019,Su2019,Mishra2020c}, an strategy for fabricating atomically precise graphene flakes on a metallic surface  \cite{Cai2010,Treier2010,Ruffieux2016,clair2019}. Despite such progresses in fabrication, their magnetic properties on  a surface have not been demonstrated experimentally.

\begin{figure}[b!]
\centering
\includegraphics[width=0.45\textwidth]{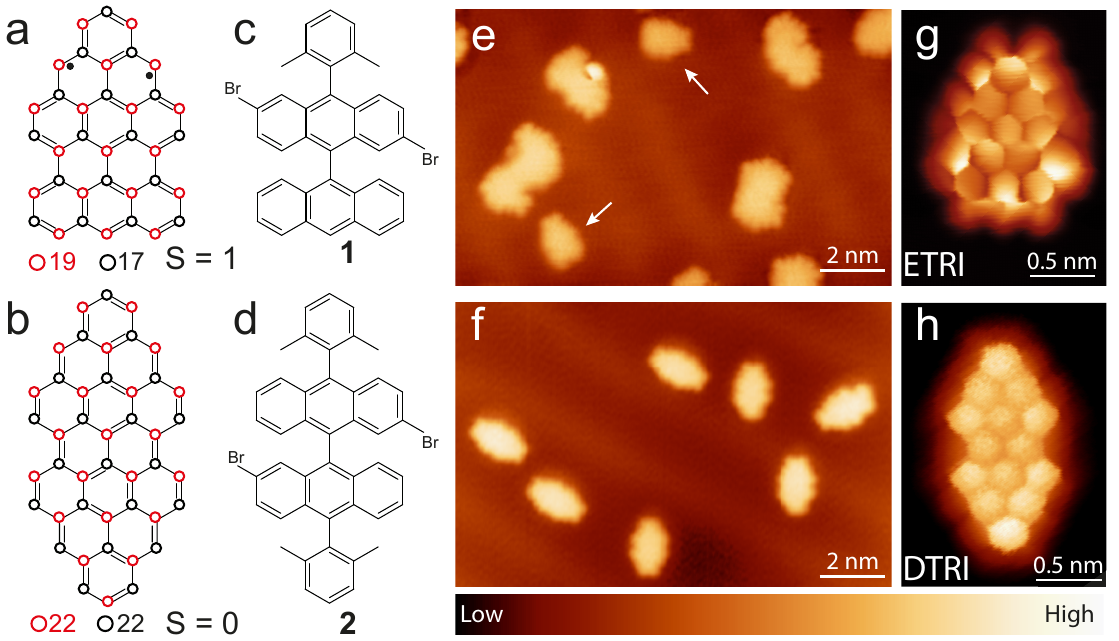}
\caption{\textrm{Chemical structures and synthetic route of extended-triangulene and double-triangulene}. \textbf{a,b}, atomic structures of ETRI and DTRI with total spin 1 and 0, respectively. Red(black) open circles denote carbon atoms belong to different sub-lattices, with the total numbers of carbon atoms indicated under the structures. \textbf{c,d}, molecular precursors 2,6-dibromo-10-(2,6-dimethylphenyl)-9,9'-bianthracene (precursor \textbf{1}) and 2,2'-dibromo-10,10'-bis(2,6-dimethylphenyl)-9,9'-bianthracene (precursor \textbf{2}) used to synthesize ETRI and DTRI, respectively. \textbf{e,f}, STM overview images (V = 0.3 V, I = 0.1nA) of the formed GNFs from precursor \textbf{1} and \textbf{2}, respectively. The arrows in \textbf{e} indicates the ETRI monomers created. \textbf{g,h}, constant height current images (V = 2 mV) of ETRI and DTRI with a CO-terminated tip. }
\end{figure}

\begin{figure*}[t!]
\centering
\includegraphics[width=0.99\textwidth]{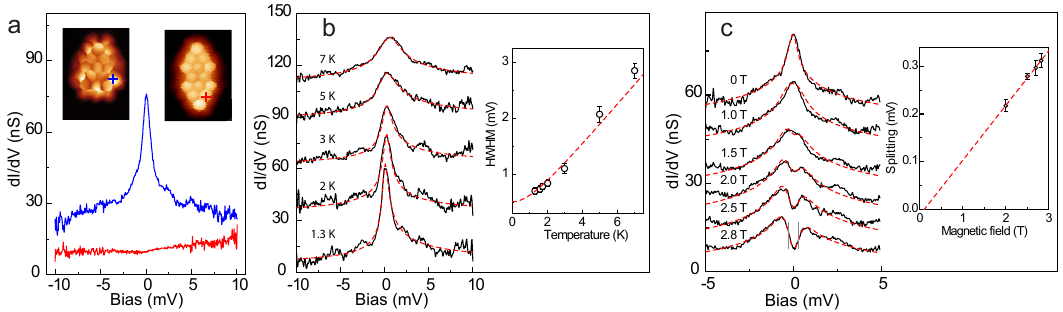}
\caption{\textrm{Kondo resonances in extended-triangulene.} \textbf{a} $dI/dV$ spectra of ETRI (blue curve) and DTRI (red curve) measured at the sites indicated by colour crosses in the inset. \textbf{b}, temperature dependence of the Kondo resonance. The half width at half maximum (HWHM) at each temperature is extracted by fitting a Frota function (red dashed lines) \cite{Frota1992}, and then corrected for the thermal broadening of the tip \cite{Zhang2013}. The inset shows the temperature dependence of HWHM, including the plot of the empirical expression $\sqrt{(\alpha k_BT)^2+(2k_BT_K)^2}$ \cite{Nagaoka2002}, \rev{that fits the temperature evolution of the HWHM with a Kondo temperature $T_K \sim 6$ K and $\alpha = 8.5$. The temperature evolution of the zero-bias conductance, shown as supplementary material \cite{SI}, agrees with a spin-polarized state in the Kondo regime with energy scale of a few Kelvin. }
\textbf{c}, magnetic field dependence of the Kondo resonance with the field strength indicated in the figure, measured at $T=1.3$ K. The red dashed lines shows the simulated curves using a model for a spin-1 system \rev{using the code of Ref.~\cite{Ternes2015}}. The inset shows the dependence of splitting on magnetic fields. The splitting $\Delta(B)$ was determined using the steepest slope, as indicated on the spectrum at 2.8 T for a example. The fitting to the splitting for a spin-1 system with equation $\Delta(B)=g\mu_Bm_sB$ gives a $g$-factor of 1.98$\pm$0.07. The spectra in \textbf{d,e} are shifted vertically for clarity. }
\end{figure*}

Here we report the OSS of triangulene-like GNFs and \rev{demonstrate that they have} a triplet ground state.   The synthesized GNFs have reduced symmetry compared to triangulene, which increases the localization of the magnetic moments. High-resolution STM images and spectroscopy allow us to identify the two spin centers on the GNFs and map their localization. Their ferromagnetic correlations are characterized by the spin-1 Kondo effect \cite{Sasaki2000,Roch2009,Parks2010}, which happens due to the incomplete screening of two coupled spins by conducting electrons of the metal substrate. Our results also show that the   strength of correlations between spins  depends on the distance between them. 

Figure 1\textbf{a} shows the chemical structures of the  GNF  characterized in the experiments, named here extended-triangulene (ETRI). This GNF has 19 carbon atoms in one sub-lattice (highlighted by red circles) and 17 carbon atoms in the other  (black circles), which results in a total spin $S=1$ according to Lieb's theorem \cite{Lieb1989}. 
For comparison, we also produced the GNF in Fig. 1\textbf{b} (named as double-triangulene (DTRI)), which has 22 carbon atoms in both sub-lattices and,  thus, a spin $S=0$ ground state. These GNFs we synthesized  by depositing the respective molecular precursors, shown in Fig.~1\textbf{c,d} (details in supplementary material (SM) \cite{SI}), on a  Au(111) surface at 330 $^{\circ}$C. The hot surface activates a simultaneous debromination and cyclodehydogenation step that planarizes them into their final structures. 

\rev{Low-temperature STM overview images of the Au (111) surface after deposition of precursors \textbf{1} and \textbf{2}, are shown in Figs.~1\textbf{e} and 1\textbf{f}, respectively. While the reacted precursor \textbf{1}, ETRI, appear on the surface mostly as monomers and dimers in a ratio of approximately 1:7, nearly all deposited precursors \textbf{2}, DTRI, remain as monomer species. In every case, the monomers adopt a  planar configuration, as expected  for the the structures in Fig.~1\textbf{a,b}. Furthermore,  high-resolution STM current images using a CO-terminated tip  \cite{Gross2009,Kichin2011}  (Fig.~1\textbf{g,h}) reproduce the chemical bond structures of Fig.~1\textbf{a,b}, indicating the successful synthesis of ETRI and DTRI. 
It is worth noting that the current image of DTRI shows merely the backbone structure, as also observed in similar symmetric systems \cite{Beyer2019}}, while the current image of ETRI shows additional bright features at the edges. Considering that these images were recorded at 2 mV, the bright features surrounding the backbone structure indicate an enhancement of the local density of states (LDOS) close to Fermi level for the ETRI molecule.

The  different shape in the images is better manifested in differential conductance ($dI/dV$) spectra measured on both types of GNFs (Fig.~2\textbf{a}). The spectrum on ETRI shows a pronounced and narrow (FWHM $\sim$ 1 meV) $dI/dV$ peak centered at zero bias. \rev{The zero-bias peak broadens anomalously fast with temperature (as described in } Fig.~2\textbf{b}) and splits with magnetic fields (Fig.~2\textbf{c}), demonstrating that it is a manifestation of the Kondo effect  \cite{Ternes2009}. A (zero-bias) Kondo-derived resonance reflects the screening of \rev{a} local spin by conduction electrons \cite{Ternes2015} and, hence, is a direct proof of the presence of localized magnetic moments on ETRI even when it lays on a metal surface. On the contrary, the spectra taken on  DTRI is featureless. Furthermore, both wide bias range spectra and $dI/dV$ maps reveal their closed-shell ground state (see SM).

\begin{figure*}[t!]
    \centering
	\includegraphics[width=0.9\textwidth]{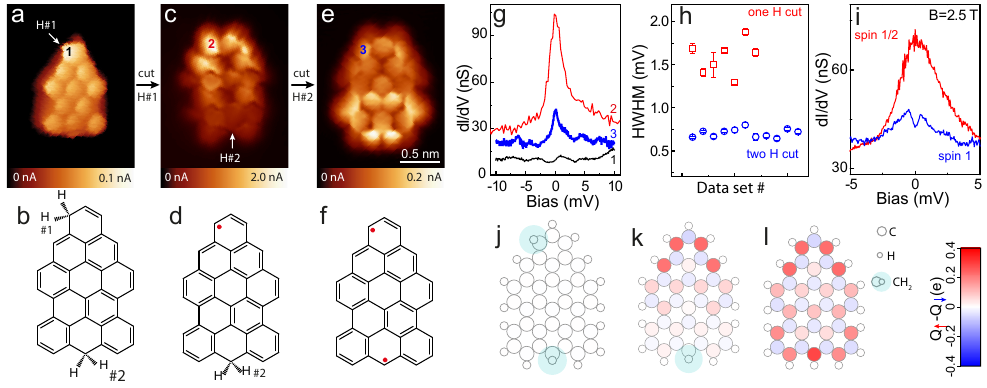}
	\caption{\textrm{Spin manipulation by electron-induced removal of extra H-atoms}. \textbf{a}, constant height current image ($V = 2$ mV) of a 2H-ETRI, with its chemical structure shown in \textbf{b}.  \textbf{c,e} constant height current image ($V = 2$ mV) of the same ETRI of \textbf{a} after the electron-induced removal of one and two passivating hydrogen atoms,respectively. The corresponding chemical structures are show in \textbf{d,f}. \textbf{g}, comparison of $dI/dV$ spectra taken on the three structures, at the locations noted with numbers in \textbf{a,c,e}. \textbf{h}, comparison of the HWHM from the Kondo resonance taken on 1H-ETRI (red data points) and ETRI (blue data points), respectively. The value of each data point is extracted by fitting the corresponding Kondo resonances with Frota functions \cite{Frota1992}. Different data points stands for the measurements from different molecules. 
	\textbf{i} comparison of Kondo resonance taken on 1H-passivated ETRI (red curve) and ETRI (blue curve) at a magnetic field of 2.5 T. \textbf{j,k,i}, Spin polarization of 2H-ETRI, 1H-ETRI and ETRI from meanfield Hubbard simulations. 
    }
\end{figure*}

The temperature and magnetic field dependence of the Kondo resonances provides further insight on the nature of the spin state of ETRI. Although the Kondo effect is frequently described on $S=1/2$ systems, it also occurs for higher spin configurations  \cite{Roch2009,Parks2010}. 
For $S=1$, a zero bias resonance reflects a partly screened spin, i.e. with only one interaction channel with the substrate  \cite{Pustilnik2001}. Similar to the spin-1/2 case, the resonance broadens with temperature (with a Kondo temperature $T_K\sim6$~K, after Fig.~2\textbf{b}, see SM), but shows a larger sensitivity to the magnetic field  \cite{Roch2009,Parks2010}. 
\rev{While the Kondo resonance of a spin-1/2 system in the strong coupling regime ($T<T_K $) splits linearly with magnetic fields only above a critical field   $B_c \ge 0.5k_BT_K/g\mu_b$ \cite{Costi2000,Li2019}, an underscreened $S=1$ conserves some magnetic moment, and its zero bias state splits already with $B>0$   \cite{Parks2010,Roch2009}. In tunneling spectra, 
such peak split should become visible as soon as the Zeeman energy is greater than the thermal broadening ($k_BT$), this is above one Tesla at the 1.2 K of our experimental setup. If the Kondo resonance in Fig.~2\textbf{b} were caused by a $S=1/2$ state, it should appear split in the spectra only for magnetic fields above 3.5 T. However, the peak appears split already at $B=1.5$ T, proving that the ETRI molecule has a  $S=1$ spin in an underscreened Kondo state on the gold substrate \cite{SI}. } 

The spin-1 configuration of the triangular  GNFs was further supported by tip-induced manipulation experiments. Due to their bi-radical character, the zigzag sites show some reactivity, and are frequently found passivated by hydrogen atoms produced during the OSS reactions  \cite{Talirz2013,Li2019}. The passivated carbon sites can be identified in high-resolution images by their larger bond length  \cite{Gross2012} due to the change from $sp^2$ to $sp^3$ hybridization. The STM image in Fig.~3\textbf{a} shows the bare backbone structure of an ETRI with two H-atoms passivating the zigzag sites indicated with arrows (as in Fig.~3\textbf{b}). The corresponding $dI/dV$ spectrum is featureless around zero bias (black curve in Fig.~3\textbf{g}), explaining the absence of bright features in the STM image. 

We first cut off one of the passivated H atoms by placing the STM tip on top of site \#1 in Fig.~3\textbf{a} and ramping up the sample bias above 1.5 V  \cite{Talirz2013,Li2019}. The H removal was monitored by a sudden step in the tunnelling current  \cite{Li2019}. The STM image afterwards appeared with an enhanced LDOS signal around the \#1 site (Fig.~3\textbf{c}) and the $dI/dV$ spectrum on site \#1 presented a  pronounced zero-energy peak (red Fig.~3\textbf{g}). We identify this to a spin-1/2 Kondo resonance  \cite{Li2019}, resulting from the single radical state recovered by the removal of the extra H \#1 atom  (Fig.~3\textbf{d}). 
Following a similar process to cleave the second H-CH bond at site \#2, recovered bright current features all around the ETRI backbone, similar to the reference bi-radical structure of Fig.~1\textbf{g}. The spectrum measured again over site \#1 shows now the Kondo resonance \rev{with  smaller amplitude and  line-width, similar to the one in Fig. 2a, agreeing with its underscreened S=1 state.     
As quantified in Fig. 3h, the Kondo resonance of the doubly dehydrogenated GNFs appeared repetitively  with  line-width about FWHM $\sim$ 0.7 meV, significantly smaller than in the  singly-passivated species (Fig.~3\textbf{g,h}), hinting to their different Kondo states. }
The spin assignment of each species was further corroborated by comparing their response to a magnetic field of 2.5 T. While the intermediate $S=1/2$ specie did not present any detectable split of the Kondo resonance, the bi-radical one exhibited a clear split, as in Fig.~2\textbf{c}, demonstrating the larger spin polarization of its $S=1$ underscreened ground state.

\begin{figure}[t!]
   \centering
	\includegraphics[width=0.47\textwidth]{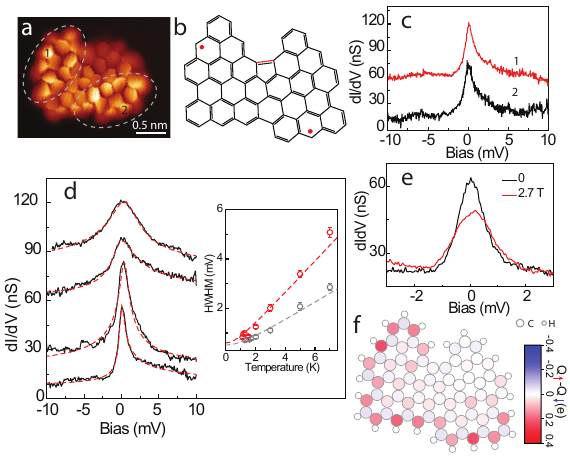}
		\caption{\textrm{Magnetic properties of dimers of extended-triangulene}. \textbf{a}, constant height current image ($V = 2$ mV) of a ETRI dimer with its chemical structure shown in \textbf{b}. \textbf{c}, $dI/dV$ spectra taken on the dimer on the sites indicated with corresponding numbers in \textbf{a}. \textbf{d}, temperature dependence of the Kondo resonance taken at location 2. The spectra are shifted vertically for clarity. The half width at half maximum (HWHM) at each temperature is extracted by fitting a Frota function (red dashed lines) \cite{Frota1992}, and then corrected for the thermal broadening of the tip \cite{Zhang2013}. The inset shows the temperature dependence of HWHM (red data points), which was simulated with the empirical expression $\sqrt{(\alpha k_BT)^2+(2k_BT_K)^2}$ \cite{Nagaoka2002}, giving a Kondo temperature $T_K \sim 7$ K with $\alpha = 15$. For comparison, the temperature dependence of HWHM from the Kondo resonance taken on a single ETRI monomer (gray data points) is also shown. \textbf{e}, Kondo resonance taken at location 2 at different magnetic fields, at $T=1.3$ K. \textbf{f}, spin polarization of a ETRI dimer from meanfield Hubbard simulations. }
\end{figure}
 
The emergence of magnetism in ETRI is reproduced by \rev{both mean field Hubbard (MFH) and density functional theory (DFT) simulations (see SM)}. Similar to the case of triangulene  \cite{Yazyev2010,Fernandez-Rossier2019}, this modified GNF present two (singly-occupied) zero-energy modes, with a triplet ground state clearly preferred over a singlet one by more than 60 meV  (see \rev{Figs.~S9 and S16}). The spin polarization (Fig.~3\textbf{l}) shows two spin centers localized at opposite sides of the triangular GNF, with a distribution that resembles the experimental current maps at zero bias (Fig.~1\textbf{g} and Fig.~3\textbf{e}). We note that their ferromagnetic exchange interaction is much larger than the Kondo energy scale ($k_B T_K< 1$ meV), thus explaining the $S=1$ underscreened Kondo ground state found in the experiments.
MFH simulations also shows that hydrogen passivation of the radical states quenches sequentially the spins, turning ETRI into a $S=1/2$ doublet (Fig.~3\textbf{k}) or completely non-magnetic (Fig.~3\textbf{j}), as demonstrated by the electron-induced removal of extra H-atoms in the experiments.

A paramagnetic ground state was also found on molecular dimers formed during the OSS process (as shown in Fig.~1\textbf{e}), but their larger size crucially affects their magnetic properties. Fig.~4\textbf{a} shows a high resolution image of a dimer. Two ETRI moieties are covalently linked together (Fig.~4\textbf{b}) following the Ullmann-like C-C coupling of their halogenated sites~ \cite{DeOteyza2016}. During the cyclodehydrogenation step, an extra pentagonal ring is created between them, as highlighted by a red bond in Fig.~4\textbf{b}, reducing the number of radicals of the dimer to only two. The bi-radical state is experimentally demonstrated in the supplementary material \cite{SI}.

The STM image of Fig.~4\textbf{a} also reproduces bright features around the dimer backbone (indicated with dashed ellipses) corresponding to the localization of the Kondo effect: $dI/dV$ spectra measured on either of the two show pronounced peaks centered at zero bias (Fig.~4\textbf{c}), which broaden with temperature (Fig.~4\textbf{d}) and magnetic field (Fig.~4\textbf{e}). However, these Kondo resonances broaden faster with $T$ than for the $S=1$ case of ETRI \rev{(both compared in the inset of Fig. 4d)}, and show no split at $B=2.7$ T, signaling a different magnetic ground state. 

Our MFH simulations of ETRI dimers like in Fig.~4\textbf{b} confirm their bi-radical state, but triplet and singlet solutions lay now closer in energy (see SM). On a surface, magnetic ground state probably behaves as two non-interacting $S=1/2$ spins. This explains the lack of $B$-induced split and the faster broadening with $T$, expected for s=1/2 Kondo systems \cite{Parks2010}. The spin polarization maps obtained from MFH simulations reproduce well the bright current regions on the dimers (Fig.~4\textbf{f}), further corroborating that this signal can be associated to the spin distribution. The origin of such small magnetic exchange between the two radical states is related to the presence of a pentagon between them, and to the larger separation between the two spin centers. In fact, in the absence of this extra C-C bond  MFH simulations  find a robust $S=2$ ground state. \rev{Thus, pentagonal rings embedded in certain sites of  an open-shell GNF can affect critically its magnetic state  by quenching a radical state and modifying the total spin (here by $\Delta$S=1) \cite{Mishra2020a}, just as extra hydrogen atoms do \cite{Li2019}, but their placement on a specific site can designed precisely during the OSS process.}

\rev{In summary, we have  demonstrated the magnetic ground state of graphene flakes fabricated deterministically with a triangular-like shape. The survival of the $S=1$ state on the metal surface is identified first through the net spin polarization of their Kondo state, which reacts to magnetic fields as an underscreened spin triplet. The $S=1$ state was further confirmed through  removal of hydrogen atoms by tip manipulation experiments, which revealed the stepwise emergence of two spins localized at different sides of the flakes.  We note that our findings here contrast with the absence of magnetic signals in previous studies of larger triangulene-flakes \cite{Mishra2019,Su2019}. It is therefore an interesting subject for future work to unveil the precise interplay between size and symmetry of GNFs for their magnetic state over a metal surface.  Nevertheless, the existence of GNFs with a net spin on a surface opens the door to new investigations of their spin dynamics and coherence  over an inorganic supports, which are crucial aspects  for utilizing graphene nanosystems \cite{Lombardi2019a} in quantum spintronics applications.}

\begin{acknowledgments}
We are indebted to Carmen Rubio for fruitful discussions. 
We acknowledge financial support from Spanish AEI 
(MAT2016-78293-C6, FIS2017-83780-P, and the Maria de Maeztu Units of Excellence Programme MDM-2016-0618), the European Union's Horizon 2020 (FET-Open project SPRING Grant.~no.~863098), the Basque Departamento de Educaci\'on through the PhD fellowship No.~PRE$\_$2019$\_$2$\_$0218 (S.S.), the Xunta de Galicia (Centro singular de investigaci\'on de Galicia 
accreditation 2016-2019, ED431G/09), and the European Regional Development Fund (ERDF). 

\end{acknowledgments}

%

\end{document}


	\title{\rev{Uncovering the triplet ground state of triangular graphene nanoflakes \\ engineered with atomic precision on a metal surface}}
	
	\author{Jingcheng Li}
	\affiliation{CIC nanoGUNE, 20018 Donostia-San Sebasti\'an, Spain}
	\author{Sofia Sanz}
	\affiliation{Donostia International Physics Center (DIPC), 20018 Donostia-San Sebasti\'an, Spain}
	
	\author{Jesus Castro-Esteban}
	\affiliation{Centro Singular de Investigaci\'on en Qu\'imica Biol\'oxica e Materiais Moleculares (CiQUS), and Departamento de Qu\'imica Org\'anica, \\ Universidade de Santiago de Compostela, 15782 Santiago de Compostela, Spain}
	
	\author{Manuel Vilas-Varela} 
	\affiliation{Centro Singular de Investigaci\'on en Qu\'imica Biol\'oxica e Materiais Moleculares (CiQUS), and Departamento de Qu\'imica Org\'anica, \\ Universidade de Santiago de Compostela, 15782 Santiago de Compostela, Spain}
	
	\author{Niklas Friedrich} 
	\affiliation{CIC nanoGUNE, 20018 Donostia-San Sebasti\'an, Spain}
	
	\author{Thomas Frederiksen}   \email{thomas_frederiksen@ehu.eus}
	\affiliation{Donostia International Physics Center (DIPC), 20018 Donostia-San Sebasti\'an, Spain}
	\affiliation{Ikerbasque, Basque Foundation for Science, 48013 Bilbao, Spain}

	\author{Diego Pe{\~{n}}a}   \email{diego.pena@usc.es}
	\affiliation{Centro Singular de Investigaci\'on en Qu\'imica Biol\'oxica e Materiais Moleculares (CiQUS), and Departamento de Qu\'imica Org\'anica, \\ Universidade de Santiago de Compostela, 15782 Santiago de Compostela, Spain}

	\author{Jose Ignacio Pascual} \email{ji.pascual@nanogune.eu}
	\affiliation{CIC nanoGUNE, 20018 Donostia-San Sebasti\'an, Spain}
	\affiliation{Ikerbasque, Basque Foundation for Science, 48013 Bilbao, Spain}


\maketitle

\tableofcontents
\clearpage
\section{Methods}
\subsection{Sample preparation and measurement details.}
	The experiments were performed on two different scanning tunneling microscopes
	(STM) operating in ultra-high vacuum, a commercial JT STM (from specs) operated
	at 1.2 K with a magnetic field up to 3 Tesla and a home made STM operating at 5 K. Both setups allow in situ
	sample preparation and transfer into the STM. The Au(111) substrate was cleaned in UHV by repeated cycles of Ne$^{+}$ ion sputtering and subsequent annealing to 730 K. The molecular precursors were sublimated from a Knudsen cell onto the clean Au(111) substrate kept at 330 $^{\circ}$C separately. The evaporation temperatures of molecular precursor \textbf{1,2} are 230 $^{\circ}$C and 210 $^{\circ}$C, respectively. Then the two samples were characterized separately. 
	A tungsten tip functionalized with a CO molecule was used for high resolution images. The high resolution images in the manuscript were acquired in constant height mode, at very small voltages, and junction resistances of typically 20 M$\Omega$. The $dI/dV$ signal was recorded using a lock-in amplifier at 760 Hz. The bias modulation is $V_\mathrm{rms}=0.1$ mV for short range spectra ($\pm$ 10 mV), and $V_\mathrm{rms}=4$ mV for wide range spectra ($\pm$ 1 V). The Kondo state presented in the  manuscript was reproducibly observed in all the species inspected that had no hydrogen passivation (more than 25). In two of them, the temperature-dependence was measured with similar results. 
	
\subsection{Simulations with the meanfield Hubbard model.}
 \rev{In order to understand} the experimental observations, we performed simulations based on the meanfield Hubbard (MFH) model, which provides a good description for carbon $\pi$-electron systems  \cite{Li2019}. \rev{Comparison with more accurate density functional theory (DFT) calculations is provided in Sec.~\ref{sec:DFT}.}
We described the $sp^2$ carbon backbone of the different molecules with the following Hamiltonian:
$
H = \sum_{ij\sigma} t_{ij} (c_{i\sigma}^\dagger c_{j\sigma}+\mathrm{h.c.})
+ U \sum_{i}\big(n_{i\uparrow}\langle n_{i\downarrow}\rangle
+\langle n_{i\uparrow}\rangle n_{i\downarrow}
-\langle n_{i\uparrow}\rangle \langle n_{i\downarrow}\rangle \big)
$
where $c_{i\sigma}^\dagger$ is the creation operator \rev{and $n_{i\sigma}=c_{i\sigma}^{\dagger}c_{i\sigma}$ the corresponding density operator} of electrons with spin $\sigma$ in the $p_z$ orbital centered at atomic site $i$.
We consider tight-binding hopping amplitudes in an orthogonal basis, $t_1=2.7$ eV, $t_2=0.2$ eV, and $t_3=0.18$ eV for first, second, and third-nearest neighbor matrix elements (defined from interatomic distances 
$d_1 < 1.6\mathrm{\AA} < d_2 < 2.6\mathrm{\AA} < d_3 < 3.1\mathrm{\AA}$)
that originate from Refs.~\cite{Li2019, Hancock2010}.
The on-site Coulomb repulsion $U=3$ eV was taken as an empirical parameter, providing a good agreement with DFT for nanographenes \cite{Li2019}, \rev{see also Sec.~\ref{sec:DFT}}.
The self-consistent solution of the MFH Hamiltonian is computed with a custom-made implementation based on \textsc{sisl} \cite{SISL}. The local spin density is obtained from the difference \rev{$\langle n_{i\uparrow}- n_{i\downarrow}\rangle$}.

\section{Synthesis of molecular precursors \textbf{1} and \textbf{2}}
\subsection{General methods for the synthesis of the precursors.}
\begin{figure}[htb!] 
\centering
	\includegraphics[width=0.5\textwidth]{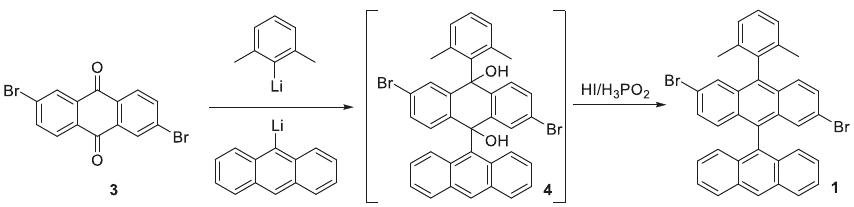}
	\caption{Synthetic route of precursor \textbf{1}.  
    }
    \label{Fig-synthesis2}
\end{figure}

\begin{figure}[htb!] 
\centering
	\includegraphics[width=0.5\textwidth]{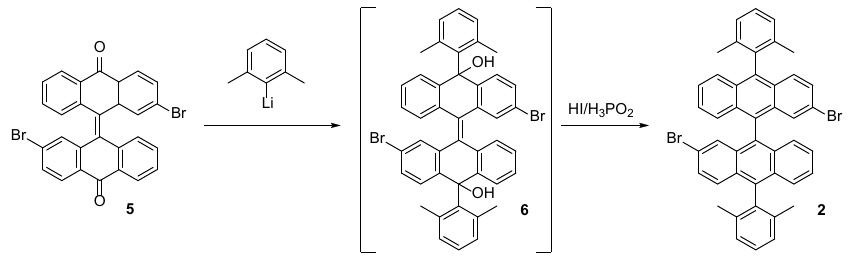}
	\caption{Synthetic route of precursor \textbf{2}.  
    }
    \label{Fig-synthesis3}
\end{figure}

All reactions were carried out under argon using oven-dried glassware. TLC was performed on Merck silica gel 60 F254; chromatograms were visualized with UV light (254 and 360 nm). Flash column chromatography was performed on Merck silica gel 60 (ASTM 230-400 mesh). 1H and 13C NMR spectra were recorded at 300 and 75 MHz or 500 and 125 MHz (Varian Mercury 300 or Bruker DPX-500 instruments), respectively. APCI spectra were determined on a Bruker Microtof instrument. The synthesis of dibromoanthrone 3 (\Figref{Fig-synthesis2}) and dibromobianthrone 5 (\Figref{Fig-synthesis3}) has been previously described \cite{Han2014,Hirshberg1953}. n-BuLi was used in solution in hexane (2.4 M). Commercial reagents were purchased from ABCR, GmbH, Aldrich Chemical Co., or Strem Chemicals Inc., and were used without further purification. Et2O was purified by a MBraun SPS -800 Solvent Purification System. 

\newpage

\subsection{Experimental details for the synthesis of precursor \textbf{1}.}

A solution of 2-iodo-1,3-dimethylbenzene (126.1 mg, 0.543 mmol) and 9-bromoanthracene (139.7 mg,0.543 mmol) in Et$_2$O (5 mL) was cooled to 0ºC. n-BuLi (452 $\mu$L, 2.4 M, 1.08 mmol) was added dropwise and the reaction mixture was stirred under argon for 1 h. The reaction was warned up slowly until room temperature, a solution of compound 3 (100.0 mg, 0.272 mmol) in Et$_2$O (20mL) was added and the mixture was stirred overnight. The reaction was quenched with 1 mL of AcOH and Et$_2$O was removed under reduced pressure. The residue was dissolved in 5 mL of AcOH, and then HI (400 $\mu$L, 3.00 mmol, 57\%) and H$_3$PO$_2$ (2.0 mL, 22.5 mmol, 50\%) were added. The reaction mixture was stirred for 2 h at 80 ºC. After cooling to room temperature, saturated aqueous solution of NaHCO$_3$ was added and the aqueous phase was extracted with CH$_2$Cl$_2$. The combined organic phases were dried over anhydrous Na$_2$SO$_4$ and the solvent was evaporated under reduced pressure. The resulting mixture was purified by column chromatography (SiO$_2$, CH$_2$Cl$_2$/hexane, 1:4) to afford 2,6-dibromo-10-(2,6-dimethylphenyl)-9,9’-bianthracene (1, 23 mg, 14$\%$) as an orange solid. $^1$CH NMR (300 MHz, CDCl$_3$) δ: 8.74 (s, 1H), 8.19 (d, J = 8.5 Hz, 2H), 7.72 (d, J =2.0 Hz, 1H), 7.54 – 7.44 (m, 3H), 7.43 (s, 1H), 7.36 (m, 3H), 7.30 – 7.27 (m, 1H), 7.25 – 7.21 (m, 2H), 7.16 (dd, J = 9.2, 2.0 Hz, 1H), 7.06 (d, J = 8.8 Hz, 2H), 6.95 (d, J = 9.3 Hz, 1H), 1.94 (s, 6H) ppm. $^{13}$CC NMR (75 MHz, CDCl$_3$) δ: 137.89 (2xC), 136.66 (C), 136.53 (C), 133.02 (C), 132.73 (C), 131.82 (2xC), 131.80 (2xC), 131.73 (C), 130.77 (C), 130.61 (C), 130.14 (3xCH), 129.37 (CH), 129.10 (CH), 129.02 (2xCH), 128.61 (CH), 128.40 (CH), 128.21 (CH), 128.15 (2xCH), 126.62 (2xCH), 126.54 (2xCH), 125.73 (2xCH), 125.59 (C), 121.18 (C), 121.11 (C), 20.34 (2xCH$_3$) ppm. MS (APCI (M+1)) for C$_{36}$H$_{25}$Br$_2$, found: 615.0318.

\subsection{Experimental details for the synthesis of precursor \textbf{2}.}

A solution of (2,6-dimethylphenyl)lithium (5.0 mL, 0.19 M in Et$_2$O, 0.93 mmol) was added dropwise to a suspension of compound 5 (50 mg, 0.093 mmol) in Et$_2$O (15 mL) and the reaction mixture was stirred at room temperature under argon for 20 h. The reaction was quenched with 1 mL of AcOH and Et$_2$O was removed under reduced pressure. The residue was dissolved in 10 mL of AcOH, and then HI (500 $\mu$L, 3.80 mmol, 57$\%$) and H$_3$PO$_2$ (1.0 mL, 9.1 mmol, 50$\%$) were added. The reaction mixture was stirred for 2 h at 80 ºC. After cooling to room temperature, saturated aqueous solution of NaHCO$_3$ was added and the aqueous phase was extracted with CH$_2$Cl$_2$. The combined organic phases were dried over anhydrous Na$_2$SO$_4$ and the solvent was evaporated under reduced pressure. The resulting mixture was purified by column chromatography (SiO$_2$, CH$_2$Cl$_2$/hexane, 1:4) to afford compound 2 (53 mg, 79$\%$) as a violet solid (mp: 180 ºC). $^1$H NMR (300 MHz, CDCl$_3$) δ: 7.62 (d, J = 8.7 Hz, 2H), 7.52 (d, J = 9.2 Hz, 2H), 7.47 (d, J = 7.0 Hz, 2H), 7.42 (d, J = 7.8 Hz, 2H), 7.40 – 7.33 (m, 6H), 7.30 (d, J = 1.9 Hz, 2H), 7.27 – 7.15 (m, 2H), 7.12 (d, J = 8.8 Hz, 2H), 1.97 (s, 6H), 1.96 (s, 6H) ppm. $^{13}$C NMR (75 MHz, CDCl$_3$) δ: 137.8 (2C), 137.6 (2C), 137.4 (2C), 137.1 (2C), 132.4 (2C), 132.1 (2C), 131.4 (2C), 129.7 (2C), 129.3 (2CH), 128.6 (2CH), 128.3 (2CH), 128.1 (2CH), 127.8 (2CH), 127.6 (2CH), 127.0 (2CH), 126.7 (2CH), 126.4 (2CH), 126.1 (2CH), 120.8 (2C), 20.3 (2CH$_3$) ppm. MS (EI) m/z ($\%$): 718 (M+, 47), 360 (5). HRMS (EI) for C$_{44}$H$_{32}$Br$_2$: calculated: 718.0871;
found: 718.0874.

\subsection{$^1$H and $^{13}$C NMR spectra of precursor \textbf{1,2}}
\begin{figure}[b!] 
\centering
	\includegraphics[width=0.99\textwidth]{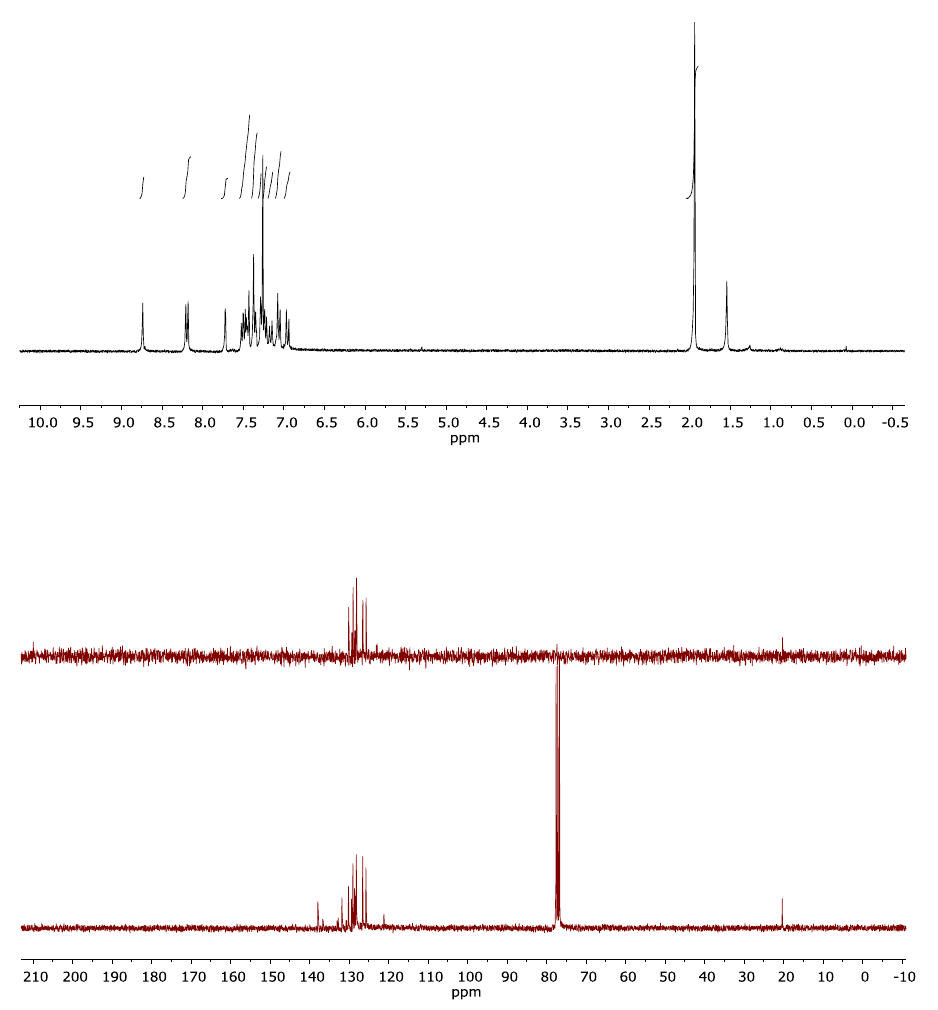}
	\caption{$^1$H and $^{13}$C NMR spectra of precursor \textbf{1}.  
    }
\end{figure}

\begin{figure}[h!] 
\centering
	\includegraphics[width=0.99\textwidth]{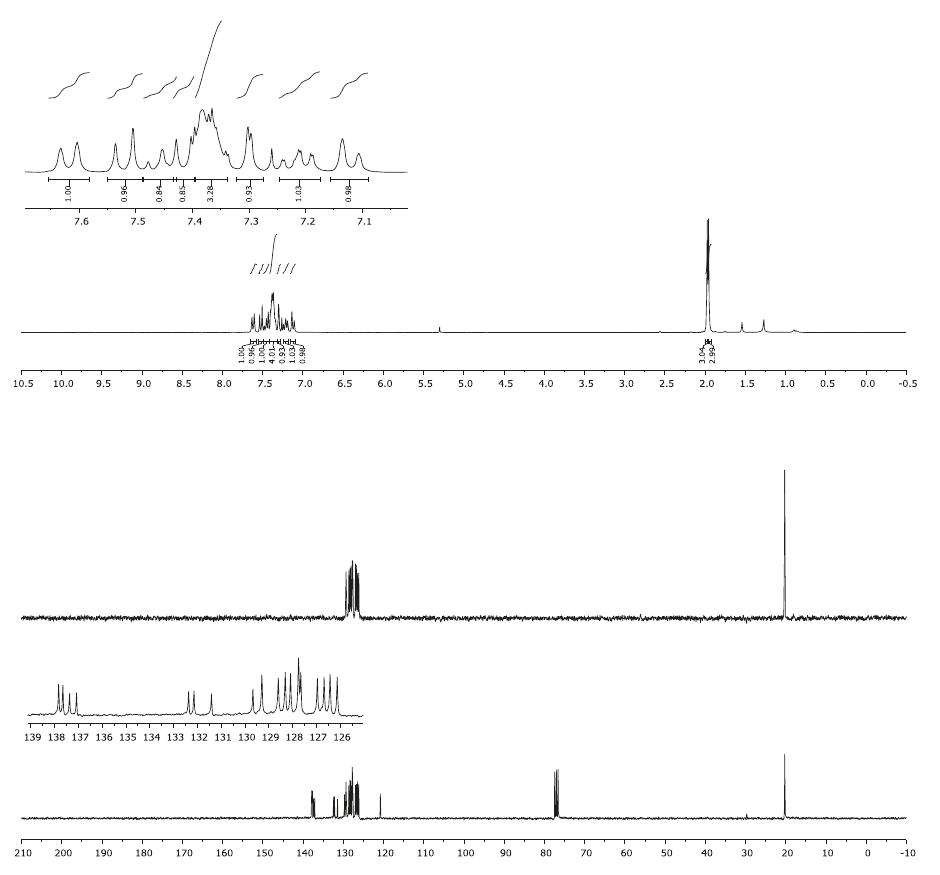}
	\caption{$^1$H and $^{13}$C NMR spectra of precursor \textbf{2}.  
    }
\end{figure}

\clearpage

\section{Temperature dependence of the zero-bias conductance}
The zero-bias conductance (equilibrium conductance) decreases with temperature following an universal form that depends on the spin. For spin 1/2, a phenomenological expression describing this behaviour was provided by Goldhaber Gordon in ref.~\cite{Goldhaber-Gordon1998a}. In ref. \cite{Parks2010}, supplementary material, this expression was further scaled to account for higher spin states. The resulting expression for the temperature dependence of the zero-bias conductance of a spin $S$ was:
\begin{equation}
G_S(T)=G(0)[(1-\Delta_S)g_S(T/T_K)+\Delta_S/2]
\end{equation}

with 
\begin{equation}
g_S(T/T_K)= [1+(T/T_K^{'})^{\xi_S}]^{-\alpha_S}
\end{equation}
and 
\begin{equation}
T_K^{'}=T_K/(2^{1/\alpha_s}-1)^{1/\xi_S}
\end{equation}

In ref.  \cite{Parks2010} these expressions were used to fit the results from  Numerical Renomalization Group simulations of various spin systems $S$ and to obtain phenomenological parameters $\Delta_S, \xi_S, \alpha_S$ for different spin states.

Here, we used eqs. 1-3 to fit the temperature dependence of the amplitude of zero-bias anomalies in ETRI, obtained for the set of spectra with HWHM values shown in  Fig.~2b of the manuscript. 
To properly obtain the resonance's amplitude, it is crucial to account for the thermal broadening of the STM's tip density of states. Therefore, we  fitted the spectral zero-bias peaks with a convolution of a Frota function times the derivative of the Fermi Dirac distribution at the corresponding temperature. 

\begin{figure}[htb!] 
	\centering
	\includegraphics[width=0.6\textwidth]{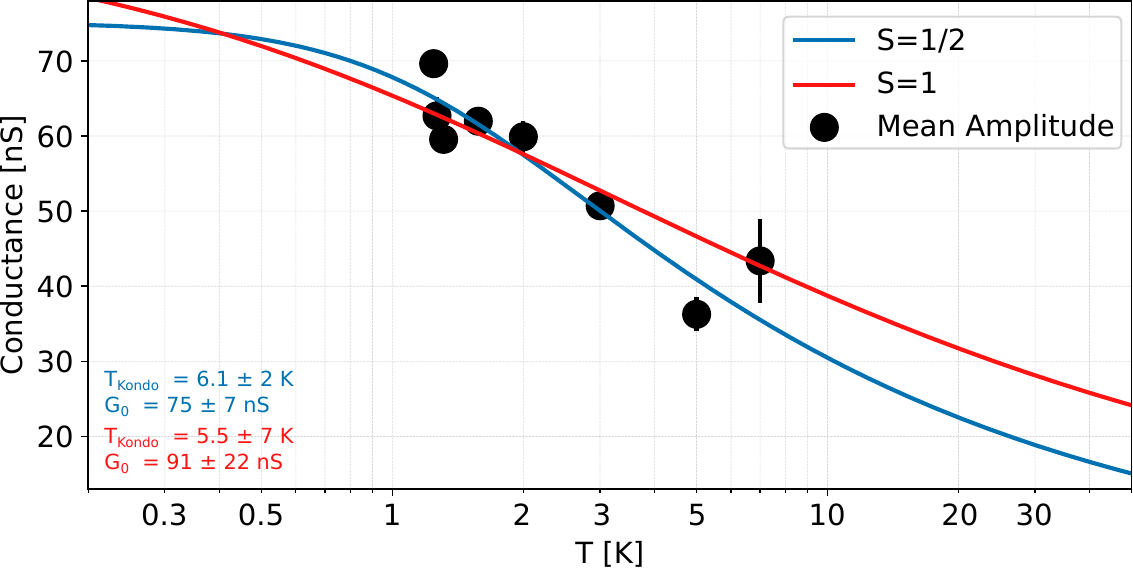}
	\caption{Temperature evolution of the zero-bias differential conductance of the Kondo resonances, and fits to the data assuming a S=1/2 (blue) and S=1 (red).    
	}		\label{KondoT}
\end{figure}

The results are shown as points in \Figref{KondoT}, including the fit of their evolution with  temperature  using the phenomenological expression from above with  $\alpha_{1/2}$=0.22 and $\alpha_{1}$=0.5, and other parameters indicated in Ref. \cite{Parks2010}  for $S=1/2$ and $S=1$ (underscreened), and only $T_K$ and $G(0)$ as fitting parameters. While the HWHM was  reasonably extracted for all spectra at the same temperature, the value $G(V=0,T)$ show more dispersion. We note that the zero bias conductance is sensitive to small variations in tip position (both vertically and laterally), while the FWHM is less affected by these. The fit is not conclusive in a specific spin state due to the data dispersion, and could be similarly fitted by either parameters of $S=1/2$ or $S=1$.  However, for $S=1/2$ a Kondo temperature of  6 K is obtained, which is a robust approximation \cite{Daroca2018}, and similar to the Kondo scale deduced from the HWHM vs. $T$ measurements of Fig.~2 \rev{of the main text}. A spin-1/2 Kondo state with Kondo temperature of 6 K is incompatible with the split of its zero-bias resonance with a magnetic field of only 1.5 T at 1.2 K, as we observed in the experiment. Following Ref.~\cite{Costi2000}, a spin 1/2 with $T_K=6$ K would split only for magnetic fields larger than 2.5 T and be detected at 1.2 K not before reaching 3.5 T, above our experimental access. This corroborates that a $S=1$ is the only possible magnetic state compatible with our experiments.

\newpage

\section{Characterization of the electronic properties of DTRI}
\begin{figure}[htb!] 
\centering
	\includegraphics[width=0.4\textwidth]{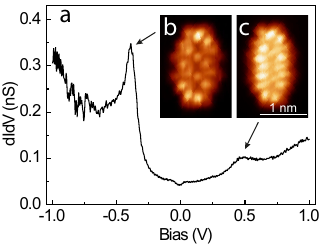}
	\caption{\textbf{Differential conductance spectrum and maps of DTRI}. Wide range $dI/dV$ spectra taken on DTRI ($V = 1.0$ V, $I = 0.1$ nA, $V_\mathrm{rms} = 0.4$ mV). The inset shows the constant height $dI/dV$ maps taken at the peak energies as indicated with arrows. The two inset images share the same scale bar.  
    }
    \label{figs1}
\end{figure}

In the main text we show that $dI/dV$ spectra taken on DTRI shows no features close to Fermi level. \Figref{figs1} here shows a wide range $dI/dV$ spectrum on DTRI. Two electronic states at -0.4 V and at 0.5 V are observed, and associated to molecular orbitals of the closed-shell DTRI. $dI/dV$ maps recorded at these energies (inset images in \Figref{figs1}) show states populating the whole DTRI, rather than the edges, and they have clearly different shapes and symmetries. The energy of the states and their spatial distributions indicate that DTRI lies in closed shell configuration.  

\section{Statistics of hydrogen passivation on ETRI.}

\begin{figure}[htb!] 
	\centering
	\includegraphics[width=0.4\textwidth]{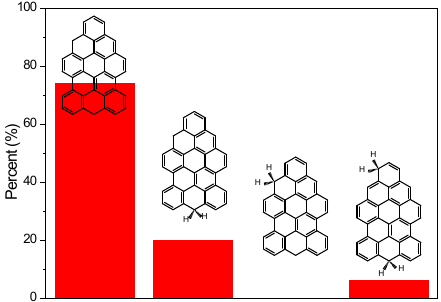}
	\caption{\textbf{Statistics of Hydrogen passivation on ETRI}. The bar plot shows the percentage of ETRI without and with hydrogen passivation found in the experiment, with their corresponding chemical structures indicated in the plot. 
	}
	\label{figs2}
\end{figure}

In the main text, we have shown that certain radicals on ETRI could be passivated by Hydrogen atoms (H-atom). Here we show the statistics of  H-atom passivation found ETRI molecules (\Figref{figs2}). From over 35 ETRI molecules characterized in the experiments, around 74\% of them shows no Hydrogen passivation, while around 20\% (6\%) of them are passivated by 1(2) hydrogen, in agreement with values found in previous experiments \cite{Talirz2013,Li2019}.

\section{Single-particle wave functions of ETRI}

\begin{figure}[h!] 
\centering
	\includegraphics[width=0.8\textwidth]{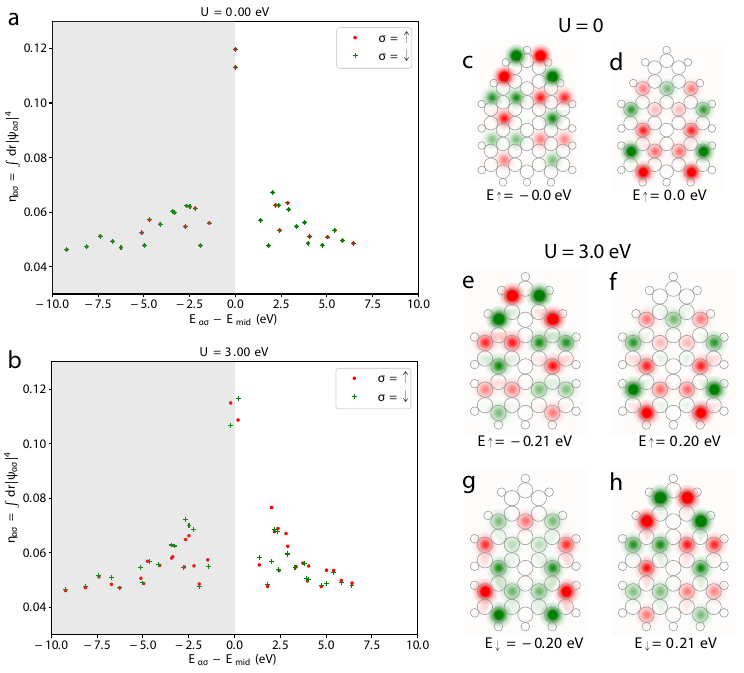}
	\caption{\textbf{Single-particle wave functions of ETRI}. \textbf{a,b}, Single-particle orbital localization $\eta_{\alpha\sigma}$ versus single-particle energy $E_{\alpha\sigma}$ in the third-nearest neighbor hopping model with two characteristic values of $U$ as noted in the figure. \textbf{c,d}, spatial distribution of wave functions of the up and down electrons (they are degenerate) with $U = 0$. \textbf{e-h}, spatial distribution of \rev{the} wave functions \rev{closest to the Fermi level for} up and down electrons \rev{in the ground state configuration ($S=1$),} with $U = 3.0$ eV.
	The single-particle energies relative to the midgap are stated below each plot.
    The size and color of the red-green circles reflect magnitude and phase of the wave function coefficients on each carbon atom, respectively.
    }
    \label{figs3}
\end{figure}

 Here we analyze the eigenspectrum of energies and states in both the noninteracting ($U=0$) and interacting ($U=3.0$ eV) MFH Hamiltonians. The degree of spatial localization of each state is computed as $\eta_{\alpha\sigma}=\int dr |\psi_{\alpha\sigma}|^4$, also denoted the inverse participation ratio \cite{Ortiz2018}. Two very localized states appear at \rev{zero energy} with $U=0$. The wave functions, shown in \Figref{figs3}\textbf{c,d} for $U=0$ and in \Figref{figs3}\textbf{e-h} for $U=3.0$ eV, are concentrated around the radical sites of the structure. 

\clearpage
\section{Ground states of ETRI from MFH simulations}

\begin{figure}[h!] 
\centering
	\includegraphics[width=0.4\textwidth]{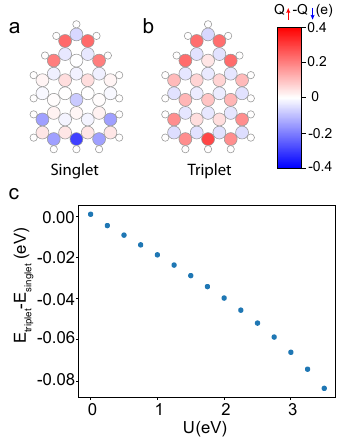}
	\caption{\textbf{Ground states of ETRI from MFH simulations}. \textbf{a,b}, Spin polarization of singlet and triplet configurations with $U = 3.0$ eV in the third-nearest neighbor hopping MFH model. \textbf{c}, energy differences between triplet and singlet ground states as a function of $U$.
    }
    \label{figs4}
\end{figure}

In the main text we show that the ETRI has spin-1 ground states. Here we analyze the ground states of ETRI from MFH simulations. With all the $U$ values considered here, the triplet configurations are always preferred \rev{energetically over the singlet configuration} (\Figref{figs4}). With $U = 3$ eV, the energy difference is \rev{more than} 60 mV.  

\clearpage
\section{Energetically preferred hydrogen passivation sites on ETRI}

\begin{figure}[htb!] 
\centering
	\includegraphics[width=0.9\textwidth]{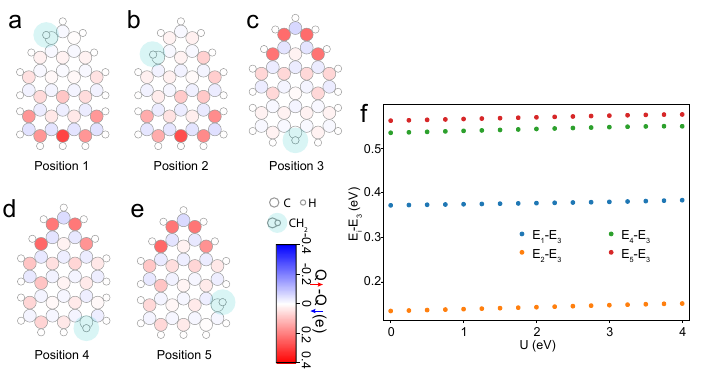}
	\caption{\textbf{Energetics and spin polarization for five energetically preferred hydrogen passivation sites from MFH simulations.} 
\textbf{a-e}, Spin polarization of the extended-triangulene molecule with H-passivated C-atom sites in five different positions from MFH simulations with $U = 3.0$ eV. \textbf{f}, comparison of total energy differences between the five different configurations for different $U$ values， with the most stable configuration (position 3) as reference.
  }
  \label{figs5}
\end{figure}

In the main text we show the hydrogen passivation on ETRI, and in \Figref{figs2} we report the statistics found for different hydrogen passivation patterns. 
To quantitatively study the relative stability of H bonded to each radical site, we computed the relative ground state energy of different hydrogen passivation schemes with MFH simulations. Figure \ref{figs5} shows that H-atom passivation at position 3 is the most stable configuration, while position 2 and 1 are the next stable configurations. The results agree with the statistics of H-atom passivation in \Figref{figs2}.

\clearpage
\section{Electron-induced removal of extra H-atoms on ETRI dimers}

\begin{figure}[htb!] 
\centering
	\includegraphics[width=0.7\textwidth]{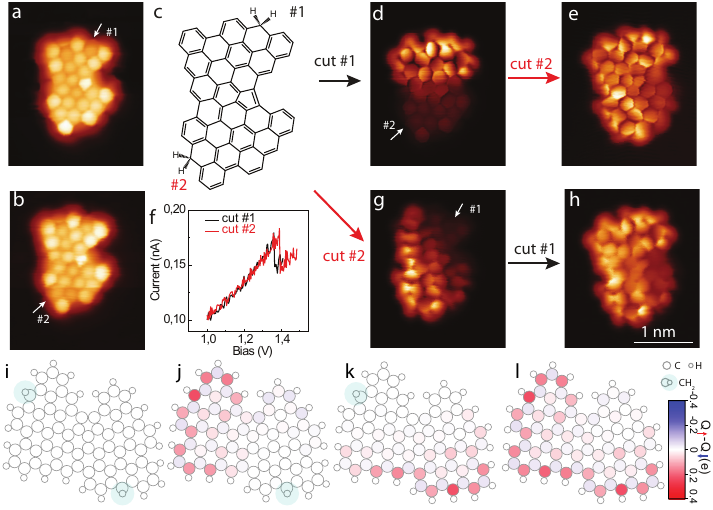}
	\caption{\textbf{Electron-induced removal of extra H-atoms on ETRI dimers}. \textbf{a,b}, constant height current images ($V=2$ mV) of two ETRI dimers with 2 hydrogen passivations each, with their chemical structures shown in \textbf{c}. \textbf{d,e}, constant height current images ($V=2$ mV) of 2H-ETRI dimer in \textbf{a} after $\#$1 and $\#$2 hydrogen atoms are step-wisely removed by tunneling electrons. The current profiles during the process of H-removal are shown in \textbf{f}. \textbf{g,h}, constant height current images ($V=2$ mV) of 2H-ETRI dimer in \textbf{b} after $\#$2 and $\#$1 hydrogen atoms are step-wisely removed by tunneling electrons. \textbf{i-l}, Spin polarization of the ETRI from our experiment with H-atom passivation, as obtained from MFH simulations with $U = 3.0$ eV.
    }
    \label{figs6}
\end{figure}

In the main text, we have shown that the electron-induced removal of extra H-atoms on 2H-ETRI proves the existence of two radicals on ETRI. Here, we show similar experiments on hydrogen passivated ETRI dimers. Figure \ref{figs6}\textbf{a,b} show high-resolution images of two dimers, each passivated by two H-atoms. Their chemical structures are shown in \Figref{figs6}\textbf{c}. 

On the ETRI dimer in \textbf{a}, we first placed the STM tip on top of extra H-atom $\#$1 and cut if off by raising the sample bias. The step-wise decrease of the tunneling current (\Figref{figs6}\textbf{f}) indicated the removal of the extra H-atom. After $\#$1 H-atom was cut off, strong LDOS appeared around the $\#$1 H-atom (\Figref{figs6}\textbf{d}). After cutting off the $\#$2 H-atom with identical procedure, the resulting STM image (\Figref{figs6}\textbf{d}) appeared with the similar current contrast as the one in Fig.~4 of the main text.

On the ETRI dimer in \textbf{b}, we first cut the $\#$2 H-atom off and found a strong LDOS emerging in the low-bias current maps now around this other site (\Figref{figs6}\textbf{g}). After subsequent removal of the $\#$1 H-atom, the image in \Figref{figs6}\textbf{h} shows nearly the same bright contrast and shape as in \Figref{figs6}\textbf{e}. 

The ETRI dimers with only 1 H-atom passivation shows very clearly the spatial distribution of the remaining unpaired electrons (\Figref{figs6}\textbf{d,g}), which reproduces the spin distribution around this region of bare ETRI dimers, as indicated by dashed gray ellipse in Fig.~4\textbf{a} in the main text. 

The experiments are further supported by MFH simulations. MFH simulations also shows 2 H-atoms passivation can completely quench the spins of ETRI dimer (\Figref{figs6}\textbf{i}), while 1 H-atom passivation turns  ETRI dimer into a $S=1/2$ doublet (\Figref{figs6}\textbf{j,k}). The spin polarizaion of 1-H atom passivated ETRI dimers from MFH simulations (\Figref{figs6}\textbf{j,k}) reproduce quite well the experimentally observed spin distribution(\Figref{figs6}\textbf{d,g}). 
\clearpage

\section{Comparison of spin-1/2 Kondo on 1 hydrogen passivated ETRI and ETRI dimer}

\begin{figure}[htb!] 
\centering
	\includegraphics[width=0.6\textwidth]{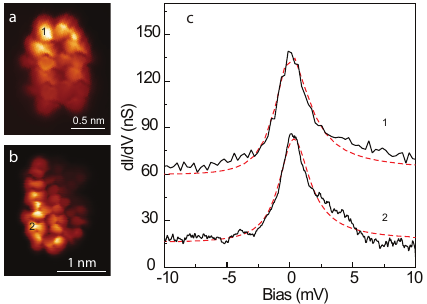}
	\caption{\textbf{Comparison of spin-1/2 Kondo on 1 hydrogen passivated ETRI and ETRI dimer}. \textbf{a,b}, constant height current images (2 mV) of 1H-ETRI and 1H-ETRI dimer. The images are the same as Fig.~3\textbf{c} and Fig. S11\textbf{g} in the main text, respectively. \textbf{c}, $dI/dV$ spectra taken on the locations as noted in \textbf{a,b}. The red dashed lines shows the fitting with Frota functions, which gives a HWHM of 1.5$\pm$ 0.03 (1.6 $\pm$ 0.05) mV for spectrum 1 (2) respectively. 
    }
    \label{figs7}
\end{figure}

In the main text we show that when the ETRI is passivated by 1 H-atom, there is only one unpaired electron left which give rise to spin-1/2 Kondo effect. In \Figref{figs6} we studied the ETRI dimer passivated by 1 H-atom. Here we compare the spin-1/2 Kondo resonance between 1H-ETRI and 1H-ETRI dimer. Figure \ref{figs7}\textbf{c} presents the Kondo resonances recorded on 1H-ETRI and 1H-ETRI dimer, which shows very similar intensity and HWHM. The results suggests similar Kondo coupling strength between unpaired electrons and electron bath for both cases.

\clearpage
\section{Single-particle wave functions of ETRI dimer}

\begin{figure}[htb!] 
\centering
	\includegraphics[width=0.8\textwidth]{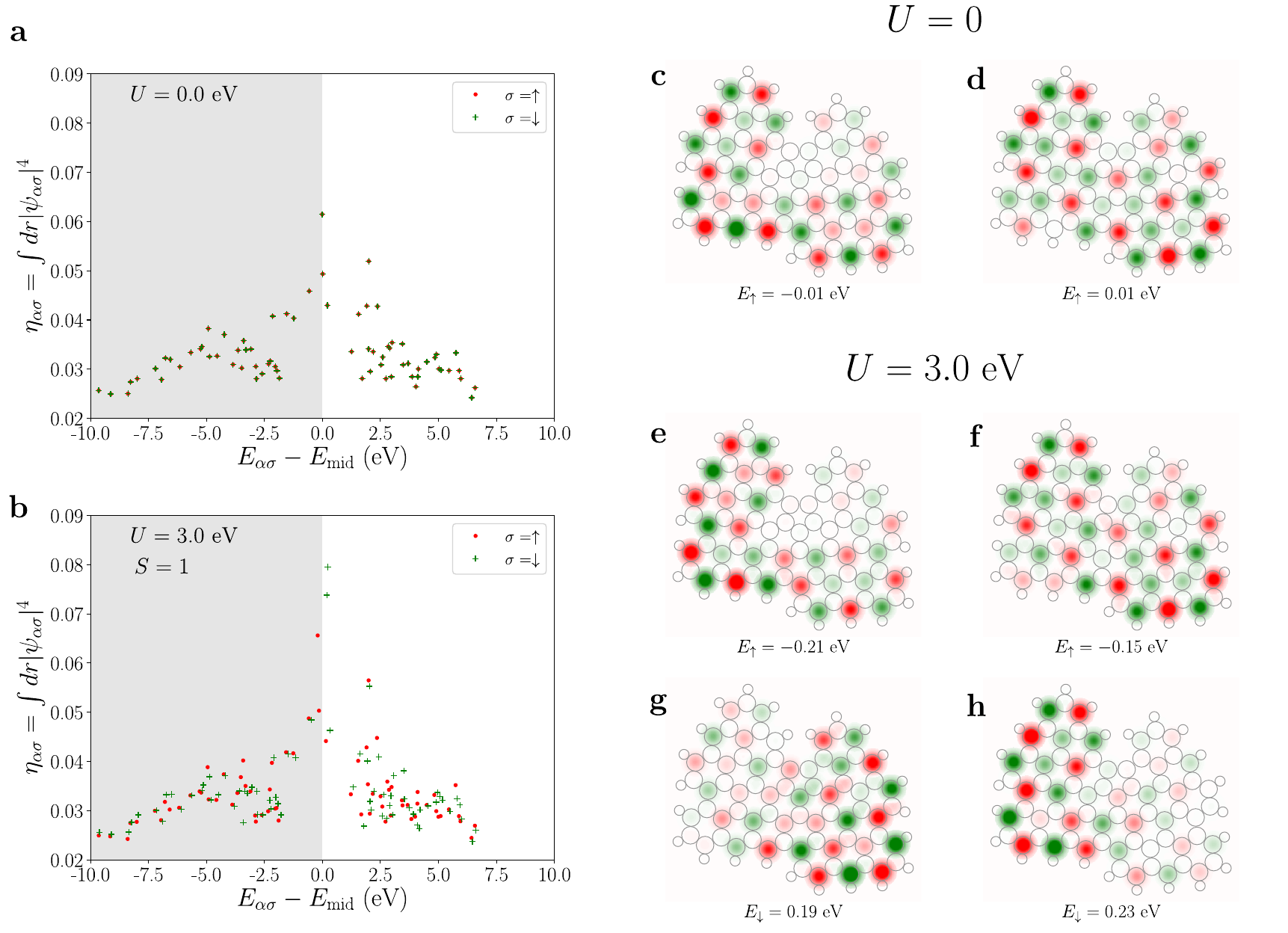}
	\caption{\textbf{Single-particle wave functions of ETRI dimer}. \textbf{a,b}, Single-particle orbital localization $\eta_{\alpha\sigma}$ versus single-particle energy $E_{\alpha\sigma}$ in the third-nearest neighbor model with two characteristic values of $U$ as noted in the figure. \textbf{c,d}, spatial distribution of wave functions of the up and down electrons(they are degenerated) with $U = 0$. \textbf{e-h}, spatial distribution of \rev{the} wave functions \rev{closest to the Fermi level for} up and down electrons \rev{in the groundstate configuration ($S=1$),} with $U = 3.0$ eV.
	The single-particle energies relative to the midgap are stated below each plot.
    The size and color of the red-green circles reflect magnitude and phase of the wave function coefficients on each carbon atom, respectively. 
    }
    \label{figs8}
\end{figure}

 Here we analyze the eigenspectrum of energies and states in both the noninteracting ($U=0$) and interacting ($U=3.0$ eV) MFH Hamiltonians for ETRI dimer. Two very localized states appear at \rev{zero energy} with $U=0$. The wave functions, shown in \Figref{figs8}\textbf{c,d} for $U=0$ and in \Figref{figs8}\textbf{e-h} for $U=3.0$ eV, are concentrated around the radical sites of the structure. 
 
\clearpage

\section{Ground state of the ETRI dimer from MFH simulations}

\begin{figure}[htb!] 
\centering
	\includegraphics[width=0.65\textwidth]{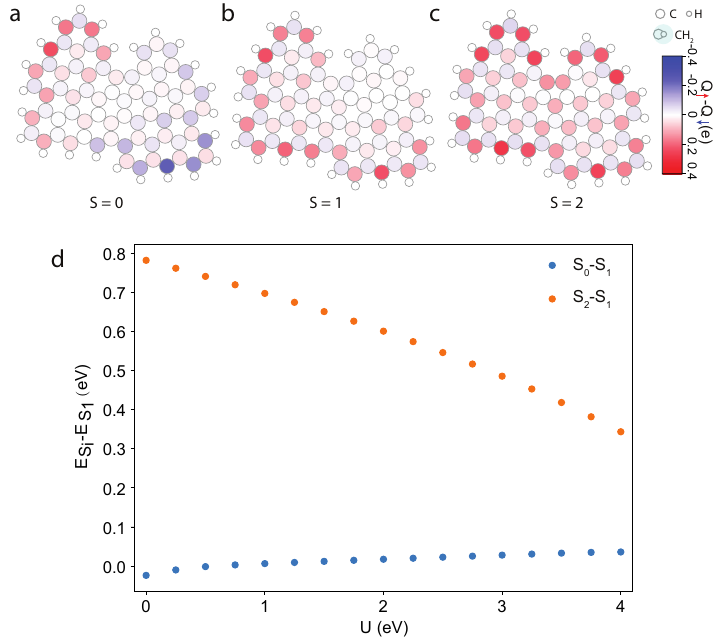}
	\caption{\textbf{Ground state of an ETRI dimer from MFH simulations}. \textbf{a,b,c}, Spin polarization of an ETRI dimer in different spin configurations with $U = 3.0$ eV from MFH simulations and \textbf{d}, their energy differences as a function of $U$.
    }
    \label{figs9}
\end{figure}

Here we analyze the ground state of the ETRI dimer from MFH simulations. With  $U$ values  larger than 0.5 eV, the triplet configuration is always preferred over the singlet state. However, the energy difference between triplet and singlet configurations is significantly reduced compared to ETRI monomers (\Figref{figs9}). This bring us to consider that spin-spin coupling on a surface is probably negligible and explains that both spins behave as two independent $S=1/2$ in the experiments. We also find that the spin-2 configurations is not preferred in energy. 

\clearpage

\section{Ground state of ETRI dimer without pentagon from MFH simulations}

\begin{figure}[htb!] 
\centering
	\includegraphics[width=0.65\textwidth]{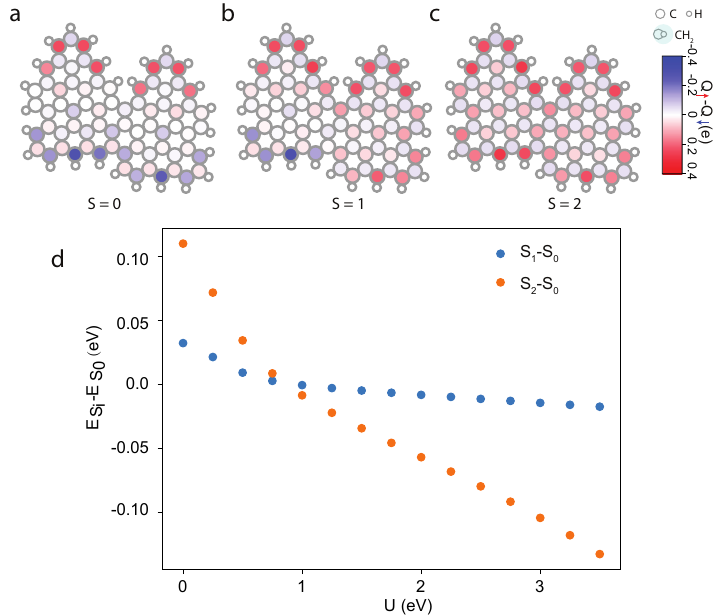}
	\caption{\textbf{Ground state of ETRI dimer without pentagon from MFH simulations}. \textbf{a,b,c}, Spin polarization of ETRI dimer without pentagon in different spin configurations with $U = 3.0$ eV from MFH simulations. \textbf{d}, energy differences between different spin configurations as a function of $U$. 
    }
    \label{figs10}
\end{figure}

In the main text and supplementary information, we showed that an extra pentagon formed in the ETRI dimer during the OSS process. The formed pentagon reduces the numbers of radicals in the dimer. Here, we study the magnetic properties of an ETRI dimer without the formation of the extra C-C bond that creates this pentagon. With the $U$ values above 1 eV, spin-2 ground states is clearly the preferred ground state (\Figref{figs10}). The comparison confirms that the formed pentagon reduces the spin density of the ETRI dimer. Hence, if the formation of this bond could be hindered during on-surface synthesis, the dimers would behave as a $S=2$ carbon paramagnets.

\clearpage

\section{Density Functional Theory simulations of ETRI and ETRI dimers}
\label{sec:DFT}
\rev{Our main simulations of the magnetic ground state and excited states are based on MFH simulations. Here, we analyze the geometry of ETRI molecules and ETRI dimers also with DFT. We performed the calculations with the SIESTA \cite{SoArGa.02.SIESTAmethodab} implementation of DFT. We used the generalized gradient approximation (GGA) \cite{PeBuEr.96.Generalizedgradientapproximation} for exchange and correlation, a 400 Ry cut off energy for the real-space grid integrations, and a double-zeta plus polarization (DZP) basis. The force tolerance was set to 2 meV/\AA\  and the density was converged to a criterion of 10$^5$.}

\begin{figure}[htb!] 
	\centering
	\includegraphics[width=0.85\textwidth]{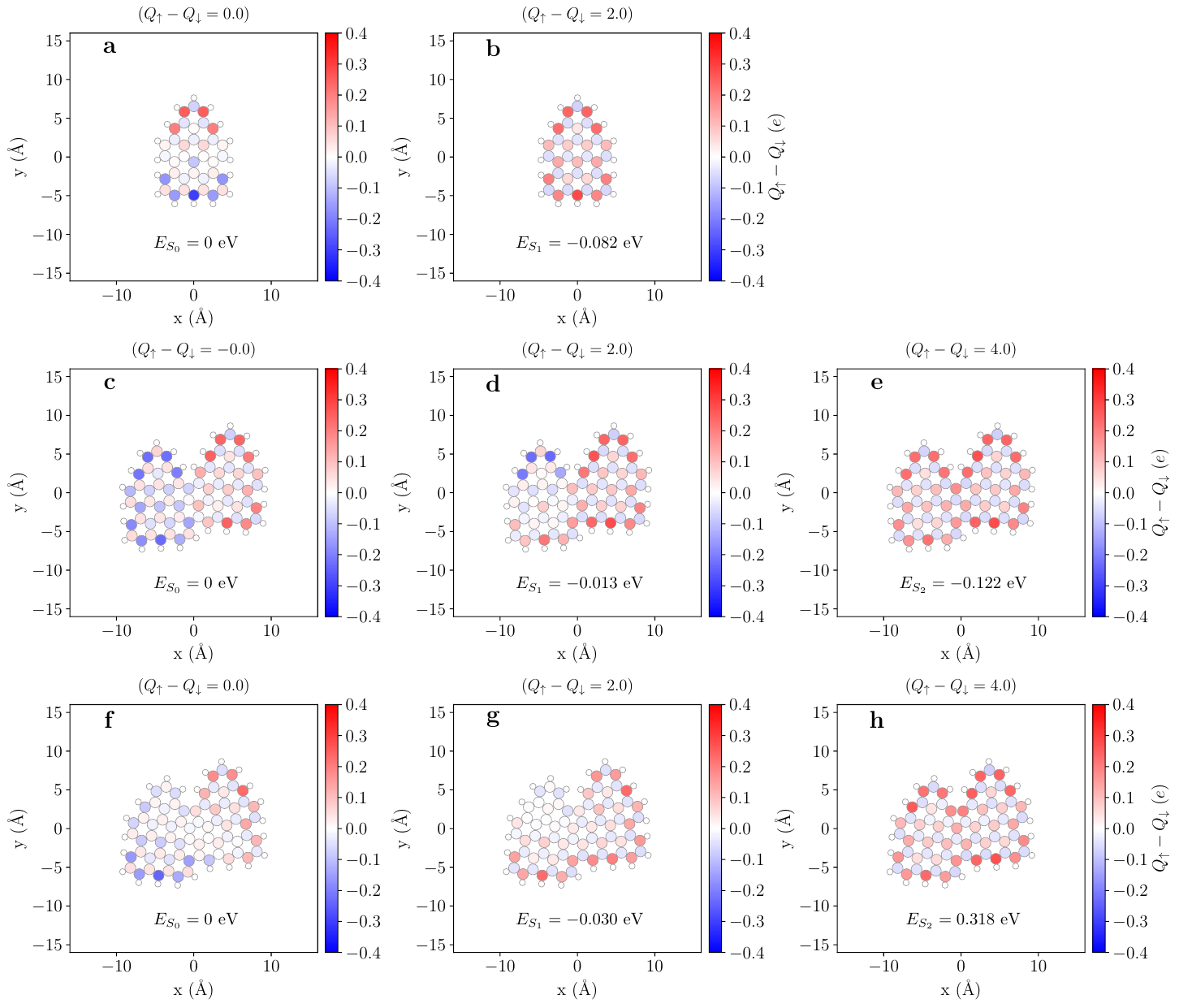}
	\caption{\textbf{Mulliken population analysis of the spin polarization obtained with DFT.}   
	(a,b) Spin polarization of the ETRI monomer with $S = 0$ ($Q_\uparrow-Q_\downarrow =0$) and $S = 1$ ($Q_\uparrow-Q_\downarrow =2e$) respectively. (c,d,e) Spin polarization of the dimer with fixed $S=0$, $S=1$, and $S=2$, respectively. (f,g,h) Spin polarization of a ETRI dimer with pentagon, imposing $S = 0$, $S = 1$ and $S = 2$, respectively. The energy of every state relative to the S=0 state is annotated in each panel.}\label{DFT}
\end{figure}

\rev{The geometries were first relaxed in the spin degenerate calculations and then frozen to a specific spin value for the spin-polarized calculations. In \Figref{DFT} we show maps the spin density obtained for different total spin-polarization imposed, namely $S = 0$ (more precisely, $Q_\uparrow-Q_\downarrow =0$), $S = 1$ ($Q_\uparrow-Q_\downarrow =2e$) and $S = 2$ ($Q_\uparrow-Q_\downarrow =4e$). The relative total energy of the different magnetic states with respect to the $S=0$ solution are annotated in the corresponding figures; a negative energy means a larger stability.  In Tab.~\ref{stability-table} we compare these results with the MFH energy values obtained with $U = 3$ eV, 3.25 eV and 3.5 eV and using the 3NN model for the hopping parameters. The results show that MFH gives an excellent approximation to the magnetic ground state of the  free flakes.   
}

\begin{table}[htb!]
\begin{tabular}{l|c|c|c|c}
& \textbf{MFH} ($U=3.0$ eV) & \textbf{MFH} ($U=3.25$ eV) & \textbf{MFH} ($U=3.5$ eV) & \textbf{DFT-GGA}\\
\hline
\hline
Monomer: $E_{S_{0}}$ & 0 & 0 & 0 & 0 \\
Monomer: $E_{S_{1}}$ &  -0.066 & -0.074 & -0.084 & -0.082\\
\hline
Dimer: $E_{S_{0}}$ & 0 & 0 & 0 & 0 \\  
Dimer: $E_{S_{1}}$ &  -0.050 & -0.057 & -0.065 & -0.013\\
Dimer: $E_{S_{2}}$ &   -0.184 &  -0.207 & -0.233 & -0.122 \\
\hline  
Dimer-pentagon: $E_{S_{0}}$ & 0 & 0 & 0 & 0 \\  
Dimer-pentagon: $E_{S_{1}}$ & -0.028 & -0.031 & -0.033 & -0.030\\
Dimer-pentagon: $E_{S_{2}}$ & 0.458 & 0.422 & 0.385 & 0.318 \\
\end{tabular}
\caption{\textbf{Comparison of relative stabilities of the different spin states from MFH and DFT}. The total energy (given in electron volts relative to the $S=0$ state) was computed for different occupations of the two spin components ($Q_\uparrow-Q_\downarrow$) within both MFH (for different $U$-values) and DFT.}
\label{stability-table}
\end{table}

\clearpage
